\documentclass[12pt,preprint,notoc,nohyper]{QCEMDA} 

\usepackage{epsfig,multicol,bbm}

\newcommand\fverb{\setbox\pippobox=\hbox\bgroup\verb}
\newcommand\fverbdo{\egroup\medskip\noindent%
            \fbox{\unhbox\pippobox}\ }
\newcommand\fverbit{\egroup\item[\fbox{\unhbox\pippobox}]}
\newbox\pippobox

\title{Hawking radiation of scalar particles from accelerating and rotating black holes}

\author{Usman A. Gillani, Mudassar Rehman and K. Saifullah  \\

Department of Mathematics, Quaid-i-Azam University, Islamabad,
Pakistan \\

Electronic address: \email{saifullah@qau.edu.pk}}

\preprint{}  

\abstract{Hawking radiation of uncharged and charged scalars from
accelerating and rotating black holes is studied. We calculate the
tunneling probabilities of these particles from the rotation and
acceleration horizons of these black holes. Using the tunneling
method we recover the correct Hawking temperature as well.}

\begin{document}

\section{Introduction}

The study of Hawking radiation from black hole as a process of
quantum tunneling has been a subject of intensive and extensive
research recently. On the one hand this approach provides physical
insight into this classically forbidden phenomenon, and on the other
it is quite rigorous mathematically. These are the reasons it has
been applied successfully to different types of configurations.
Emission of fermions has been studied for the Kerr, Kerr-Newman,
Taub-NUT, G\"{o}del, dilatonic black holes, black rings and black
holes with acceleration and rotation \cite{KM06, KM08a, KM08b, ZL,
LRW, CJZ, Ji, ZY, DJ, YY, Zha, GS, RS}. Radiation of scalars has
also been discussed in the literature \cite{BM, BKM, Ya}.

Continuing in the same spirit, in this paper, we study Hawking
radiation of scalars from a family of Pleba\'{n}ski-Demia\'{n}ski
black holes. We use the WKB approximation and Hamilton-Jacobi method
to solve the Klein-Gordon equation. We study both the neutral as
well as charged scalars. We see that the tunneling probability for
scalars is the same as that for fermions for these black holes
\cite{GS, RS}. We also recover the correct Hawking temperature
\cite{BS}.

The paper is organized as follows. In Section 2 we describe the
spacetime for accelerating and rotating black holes with electric
and magnetic charges and their basic properties. Sections 3 and 4
deal with the radiation of uncharged and charged particles
respectively, in the background of these black holes. Here tunneling
probability and Hawking temperature is calculated. In Section 5 we
study the radiation from the acceleration horizon. We conclude
briefly in Section 6.

\section{Accelerating and rotating black holes with electric and magnetic charges}

The Pleba\'{n}ski-Demia\'{n}ski metric \cite{GP05, GP06, PD} covers
a large family of spacetimes which include, among others, the
well-known black hole solutions like Schwarzschild,
Reissner-Nordstr\"{o}m, Kerr, Kerr-Newman, Kerr-NUT and many others.
Here we study a special case of this family - black holes  with
acceleration and rotation. In spherical coordinates this metric can
be written as \cite{BS, GP05, GP06, PK, DL03a, DL03b}

\begin{eqnarray}
ds^{2} &=&-\left( \frac{Q-a^{2}P\sin ^{2}\theta }{\rho ^{2}\Omega
^{2}} \right) dt^{2}+\left( \frac{\rho ^{2}}{Q\Omega ^{2}}\right)
dr^{2}+\left(
\frac{\rho ^{2}}{P\Omega ^{2}}\right) d\theta ^{2}  \nonumber \\
&&+\left( \frac{\sin ^{2}\theta \left[ P\left( r^{2}+a^{2}\right)
^{2}-a^{2}Q\sin ^{2}\theta \right] }{\rho ^{2}\Omega ^{2}}\right)
d\phi ^{2}
\nonumber \\
&&-\left( \frac{2a\sin ^{2}\theta \left[ P\left( r^{2}+a^{2}\right)
-Q\right] }{\rho ^{2}\Omega ^{2}}\right) dtd\phi .  \label{2}
\end{eqnarray}
where

\begin{eqnarray}
\Omega &=&1-\alpha r\cos \theta , \label{3.6} \\
\rho ^{2} &=&r^{2}+a^{2}\cos ^{2}\theta ,  \label{3.7} \\
P &=&1-2\alpha M\cos \theta +\left[ \alpha ^{2}\left(
a^{2}+e^{2}+g^{2}\right) \right] \cos ^{2}\theta ,  \label{3.8} \\
Q &=&\left[ \left( a^{2}+e^{2}+g^{2}\right) -2Mr+r^{2}\right] \left(
1-\alpha ^{2}r^{2}\right) .  \label{3.9}
\end{eqnarray}
Here $M$ is the mass of the black hole, $e$ and $g$ are its electric
and magnetic charges, $a$ is angular momentum per unit mass, and
$\alpha$ is the acceleration of the black hole. We use the following
notation \cite{KM08a}

\begin{eqnarray}
f\left( r,\theta \right) &=&\left( \frac{Q-a^{2}P\sin ^{2}\theta
}{\rho
^{2}\Omega ^{2}}\right) ,  \label{3.1} \\
g\left( r,\theta \right) &=&\frac{Q\Omega ^{2}}{\rho ^{2}},
\label{3.2} \\
\Sigma \left( r,\theta \right) &=&\left( \frac{\rho ^{2}}{P\Omega
^{2}}
\right) ,  \label{3.3} \\
K\left( r,\theta \right) &=&\left( \frac{\sin ^{2}\theta \left[
P\left( r^{2}+a^{2}\right) ^{2}-a^{2}Q\sin ^{2}\theta \right] }{\rho
^{2}\Omega ^{2}}
\right) ,  \label{3.4} \\
H\left( r,\theta \right) &=&\left( \frac{a\sin ^{2}\theta \left[
P\left( r^{2}+a^{2}\right) -Q\right] }{\rho ^{2}\Omega ^{2}}\right)
,   \label{3.5}
\end{eqnarray}
so that the metric (\ref{2}) can be written as

\begin{equation}
ds^{2}=-f\left( r,\theta \right) dt^{2}+\frac{dr^{2}}{g\left(
r,\theta \right) }+\Sigma \left( r,\theta \right) d\theta
^{2}+K\left( r,\theta \right) d\phi ^{2}-2H\left( r,\theta \right)
dtd\phi .  \label{3}
\end{equation}

The electromagnetic vector potential for these black holes is
\cite{GP06}
\begin{equation}
\mathbf{A}=\frac{-er\left[ dt-a\sin ^{2}\theta d\phi \right] -g\cos
\theta \left[ adt-\left( r^{2}+a^{2}\right) d\phi \right]
}{r^{2}+a^{2}\cos ^{2}\theta }. \label{4}
\end{equation}
The event horizons are obtained by putting $g(r,\theta)=0$, which
gives their location at

\begin{equation}
r=\frac{1}{\alpha \cos \theta },r=\pm \frac{1}{\alpha },\textrm{ and
} r_{\pm }=M\pm \sqrt{M^{2}-a^{2}-e^{2}-g^{2}}. \label{6}
\end{equation}
Here $r_{\pm }$ represent the outer and inner horizons corresponding
to the Kerr-Newman black holes. The other horizons are acceleration
horizons. The angular velocity at the horizon is \cite {KM08a, GS}

\begin{equation}
\Omega _{H}=\frac{a}{\left( r_{+}^{2}+a^{2}\right) }.  \label{7}
\end{equation}
We define another function which will be needed later
\[
F\left( r,\theta \right) =f\left( r,\theta \right)
+\frac{H^{2}\left( r,\theta \right) }{K\left( r,\theta \right) }.
\]
Using the values of $f$, $K$ and $H$ this takes the form
\begin{equation}
F\left( r,\theta \right) =\frac{PQ\rho ^{2}}{\Omega ^{2}\left[
P\left( r^{2}+a^{2}\right) ^{2}-a^{2}Q\sin ^{2}\theta \right] }.
\label{8}
\end{equation}

\section{Tunneling of scalar particles}

In this section we consider the uncharged black hole, that is, we
take $e=0=g$ in Eq. (\ref{2}). The Klein Gordon equation for the
scalar field $ \phi $ is
\begin{equation}
g^{\mu \nu }\partial _{\mu }\partial _{\nu }\Phi -\frac{m^{2}}{\hbar ^{2}}%
\Phi =0.  \label{33}
\end{equation}
We apply the WKB approximation and assume an ansatz of the form

\begin{equation}
\Phi \left( t,r,\theta ,\phi \right) =\exp \left[ \frac{\iota }{\hbar }%
I\left( t,r,\theta ,\phi \right) +I_{1}\left( t,r,\theta ,\phi
\right) +O\left( \hbar \right) \right] .
\end{equation}
Putting this in Eq. (\ref{33}) and simplifying gives

\begin{equation}
-[g^{\mu \nu }\partial _{\mu }I\partial _{\nu }I+m^{2}]+O(\hbar )=0.
\end{equation}
To simplify further, we consider the given metric in co-rotating
frame, $\hbar $ has been set equal to unity and the classical action
$I$ satisfies the equation

\begin{equation}
g^{\mu \nu }\partial _{\mu }I\partial _{\nu }I+m^{2}=0.
\end{equation}%
Taking variation on $\mu$, $\nu $

\begin{equation}
g^{tt}\left( \partial _{t}I\right) ^{2}+g^{rr}\left( \partial
_{r}I\right) ^{2}+g^{\theta \theta }\left( \partial _{\theta
}I\right) ^{2}+g^{\phi \phi }\left( \partial _{\phi }I\right)
^{2}+2g^{t\phi }(\partial _{t}I)(\partial _{\phi }I)+m^{2}=0.
\end{equation}%
Substituting the values from the metric (\ref{2}) this takes the
form

\begin{eqnarray}
0 &=&-\frac{\left( \partial _{t}I\right) ^{2}}{F(r,\theta
)}+g(r,\theta )\left( \partial _{r}I\right) ^{2}+\frac{\left(
\partial _{\theta }I\right)
^{2}}{f^2 (r,\theta )}+\frac{f(r,\theta )}{F(r,\theta )K(r,\theta )}%
\left( \partial _{\phi }I\right) ^{2} \nonumber \\
&&+\frac{-2H(r,\theta )}{F(r,\theta )K(r,\theta )}(\partial
_{t}I)(\partial _{\phi }I)+m^{2}. \label{3.6}
\end{eqnarray}
Taking into account the symmetries of the spacetime at hand we
assume the separation of variables in the action as

\begin{equation}
I=-Et+W\left( r\right) +J\phi . \label{ans}
\end{equation}%
Plugging into Eq. (\ref{3.6}) and evaluating at the horizon $r=r_+$
we get

\begin{equation}
0=-\frac{1}{F(r_{+},\theta )}\left( E-\frac{H(r_{+},\theta
)}{K(r_{+},\theta )}J\right) ^{2}+\frac{1}{K(r_{+},\theta
)}J^{2}+g(r_{+},\theta )(W^{^{\prime }}(r))^{2}+m^{2}.  \label{36}
\end{equation}%
Using Eqs. (\ref{3.1}) to (\ref{3.5}) and (\ref{8}) in this, we
obtain

\begin{eqnarray*}
0 &=&-\frac{\Omega ^{2}(r_{+},\theta)(r_{+}^{2}+a^{2})^{2}}{%
(r-r_{+})Q^{^{\prime }}(r_{+})\rho ^{2}(r_{+},\theta )}\left(
E-\Omega
_{H}J\right) ^{2}+\frac{1}{K(r_{+},\theta )}J^{2}+m^{2} \\
&&+\frac{(r-r_{+})Q^{^{\prime }}(r_{+})}{\rho ^{2}(r_{+},\theta
)}\Omega ^{2}(r_{+},\theta )(W^{^{\prime }}(r))^{2}.
\end{eqnarray*}%
Imposing fixed $\theta =\theta _{0}$ in order to solve for $W$

\begin{eqnarray}
W_{\pm }(r) &=&\pm \int
\frac{(r_{+}^{2}+a^{2})dr}{(r-r_{+})Q^{^{\prime
}}(r_{+})}   \nonumber \\
&&\times \sqrt{\left( E-\Omega _{H}J\right)
^{2}-\frac{(r-r_{+})Q^{^{\prime }}(r_{+})\rho ^{2}(r_{+},\theta
)}{\Omega ^{2}(r_{+},\theta)}\left(m^{2}+\frac{1}{K(r_{+},\theta
)}J^{2}\right)},
\end{eqnarray}
where $+/-$ correspond to outdoing/incoming solutions. Here
$r=r_{+}$ is a simple pole; integrating around the pole we get

\begin{eqnarray}
W_+(r) &=&\frac{\iota \pi (r_{+}^{2}+a^{2})\left( E-\Omega _{H}J\right) }{%
2(r_{+}-M)(1-\alpha ^{2}r_{+}^{2})},    \\
Im W_+(r) &=&\frac{\pi (r_{+}^{2}+a^{2})\left( E-\Omega _{H}J\right)
}{2(r_{+}-M)(1-\alpha ^{2}r_{+}^{2})}.
\end{eqnarray}%
So the tunneling probabilities scalar particles is
\begin{eqnarray}
P_{emission} \propto exp[-2 Im I] &=& exp[-2(\ln W_{+}+\ln \Theta )], \\
P_{absorption} \propto exp[-2 Im I] &=& exp[-2(\ln W_{-}+\ln \Theta
)].
\end{eqnarray}%
Since $Im W_{+}=- Im W_{-}$, the probability of a particle tunneling
from inside to outside the horizon \cite{SP, SPS}, $\Gamma \propto
P_{emission}/P_{absorption}$, takes the form
\begin{equation}
\Gamma =exp[-4\ln W_{+}].
\end{equation}%
The resulting tunneling probability is
\begin{equation}
\Gamma =exp[-2\pi \frac{(r_{+}^{2}+a^{2})\left( E-\Omega _{h}J\right) }{%
(r_{+}-M)(1-\alpha ^{2}r_{+}^{2})}],
\end{equation}%
and the correct Hawking temperature, $T_{H}=\frac{\hbar \kappa
}{2\pi }$, $\kappa$ being the surface gravity, is recovered
\cite{BS}

\begin{equation}
T_{H}=\frac{(1-\alpha r_{+})^{3}(1+\alpha
r_{+}\sqrt{M^{2}-a^{2}})}{2\pi (2M^{2}+2M\sqrt{M^{2}-a^{2}})}.
\end{equation}

\section{Charged scalar particles}

In this section we find the tunneling probability of charged scalar
particles from the accelerating and rotating charged black hole at
the outer horizon $r=r_{+}=M+\sqrt{M^{2}-a^{2}-e^{2}-g^{2}}$. The
Klein-Gordon equation for the scalar field $\Phi $ with charge $q$
is given by
\begin{equation}
g^{\mu \nu }\left( \partial _{\mu }-\frac{iq}{\hbar }A_{\mu }\right)
\left(
\partial _{\nu }-\frac{iq}{\hbar }A_{\nu }\right) \Phi -\frac{m^{2}}{\hbar
^{2}}\Phi =0,  \label{d1.1}
\end{equation}%
where $m$ is the mass of the scalar particle, $q$ is its charge,
$g^{\mu \nu }$ is the inverse of metric tensor and $A_{\mu }$ is the
vector potential which is given by Eq. (\ref{4}). Using an ansatz
similar to the one in the uncharged case, the above equation takes
the form

\begin{equation}
g^{\mu \nu }\left( \partial _{\mu }I-qA_{\mu }\right) \left(
\partial _{\nu }I-qA_{\nu }\right) +m^{2}=0.   \label{d2}
\end{equation}%

Substituting the values as in the previous case and simplifying we
get
\begin{eqnarray}
0 &=&-\frac{\left( \partial _{t}I-qA_{t}\right) ^{2}}{F\left(
r,\theta
\right) }+g\left( r,\theta \right) \left( \partial _{r}I\right) ^{2}-\frac{%
2H\left( r,\theta \right) }{F\left( r,\theta \right) K\left(
r,\theta \right) }\left( \partial _{t}I-qA_{t}\right) \left(
\partial _{\phi
}I-qA_{\phi }\right)  \nonumber \\
&&+\frac{f\left( r,\theta \right) }{F\left( r,\theta \right) K\left(
r,\theta \right) }\left( \partial _{\phi }I-qA_{\phi }\right) ^{2}+\frac{%
\left( \partial _{\theta }I\right) ^{2}}{\rho ^{2}\left( r,\theta \right) }%
+m^{2}.  \label{d3}
\end{eqnarray}%
In order to solve this, we again assume an action of the form given
in Eq. (\ref{ans}) and substitute in the above equation and evaluate
at the horizon. Near the horizon $r=r_{+}$, the functions defined in
Section 2 take the following form

\begin{eqnarray}
F\left( r,\theta \right) &=&\left( r-r_{+}\right) F_{r}\left(
r_{+},\theta \right) \nonumber \\
&=&\frac{\left( r_{+}^{2}+a^{2}\cos
^{2}\theta \right) \left( 2r_{+}-2M\right) \left( 1-\alpha
^{2}r_{+}^{2}\right) }{\left( 1-\alpha r_{+}\cos \theta \right)
^{2}\left( r_{+}^{2}+a^{2}\right) ^{2}}\left(
r-r_{+}\right) , \\
g\left( r,\theta \right) &=&\left( r-r_{+}\right) g_{r}\left(
r_{+},\theta \right) \nonumber \\
&=& \frac{\left( 1-\alpha r_{+}\cos \theta \right) ^{2}\left(
2r_{+}-2M\right) \left( 1-\alpha ^{2}r_{+}^{2}\right) }{\left(
r_{+}^{2}+a^{2}\cos ^{2}\theta \right) }\left( r-r_{+}\right) , \\
\Omega _{H} &=&\frac{H\left( r_{+},\theta \right) }{K\left(
r_{+},\theta
\right) }=\frac{a}{r_{+}^{2}+a^{2}}, \\
K\left( r_{+},\theta \right) &=&\left( \frac{P\sin ^{2}\theta \left(
r_{+}^{2}+a^{2}\right) ^{2}}{\rho ^{2}\left( r_{+},\theta \right)
\Omega
^{2}\left( r_{+},\theta \right) }\right) , \\
H\left( r_{+},\theta \right) &=&\left( \frac{aP\sin ^{2}\theta
\left( r_{+}^{2}+a^{2}\right) }{\rho ^{2}\left( r_{+},\theta \right)
\Omega ^{2}\left( r_{+},\theta \right) }\right) .
\end{eqnarray}
Using these values in Eq. (\ref{d3}) and expanding near the horizon
$r=r_{+}$ we get
\begin{eqnarray*}
0 &=&-\frac{\left( 1-\alpha r_{+}\cos \theta \right) ^{2}\left(
r_{+}^{2}+a^{2}\right) ^{2}}{2\left( r_{+}^{2}+a^{2}\cos ^{2}\theta
\right) \left( r_{+}-M\right) \left( 1-\alpha ^{2}r_{+}^{2}\right)
\left( r-r_{+}\right) }\left( E-\Omega _{H}J-\frac{qer_{+}}{\left(
r_{+}^{2}+a^{2}\right) }\right) ^{2}\nonumber  \\ &+& \frac{\left( J-qA_{\phi }\right) ^{2}}{%
K\left( r_{+},\theta \right) } +\frac{2\left( 1-\alpha r_{+}\cos
\theta \right) ^{2}\left( r_{+}-M\right) \left( 1-\alpha
^{2}r_{+}^{2}\right) \left( r-r_{+}\right) }{\left(
r_{+}^{2}+a^{2}\cos ^{2}\theta \right) }W^{\prime 2}\left( r\right)
+m^{2}. \label{d7.1}
\end{eqnarray*}
Solving this equation for $W\left( r\right) $ we get

$W_{\pm }\left( r\right) = \pm \int \frac{\left( r_{+}^{2}+a^{2}\right) }{%
2\left( r_{+}-M\right) \left( 1-\alpha ^{2}r_{+}^{2}\right) \left(
r-r_{+}\right) }dr \nonumber  \\
 \times \sqrt{\left( E-\Omega _{H}J-\frac{qer_{+}}{\left(
r_{+}^{2}+a^{2}\right) }\right) ^{2}-\frac{2\left(
r_{+}^{2}+a^{2}\cos ^{2}\theta \right) \left( 1-\alpha
^{2}r_{+}^{2}\right) \left( r_{+}-M\right) \left( r-r_{+}\right)
}{\left( 1-\alpha r_{+}\cos \theta \right) ^{2}}\left( \frac{\left(
J-qA_{\phi }\right) ^{2}}{K\left( r_{+},\theta \right)
}+m^{2}\right) }.$
\newline
Here $r=r_{+}$ is the singularity, so integrating the above equation
using the residue theory we get
\begin{equation}
W_{\pm }\left( r\right) =\pm \frac{\pi i\left( E-\Omega _{H}J-\frac{qer_{+}}{%
\left( r_{+}^{2}+a^{2}\right) }\right) \left( r_{+}^{2}+a^{2}\right) }{%
2\left( r_{+}-M\right) \left( 1-\alpha ^{2}r_{+}^{2}\right) }.
\label{d10}
\end{equation}%
or%
\begin{equation}
Im W_{\pm }\left( r\right) =\pm \frac{\pi \left( E-\Omega _{H}J-\frac{%
qer_{+}}{\left( r_{+}^{2}+a^{2}\right) }\right) \left(
r_{+}^{2}+a^{2}\right) }{2\left( r_{+}-M\right) \left( 1-\alpha
^{2}r_{+}^{2}\right) }.  \label{d11}
\end{equation}%
So the tunneling probability becomes
\begin{equation}
\Gamma =\exp \left[ -4 Im W_{+}\right] .
\end{equation}%
Using the value of $Im W_{+}$ in the above equation we get%
\begin{equation}
\Gamma =\exp \left[ -\frac{2\pi \left( r_{+}^{2}+a^{2}\right)
}{\left(
r_{+}-M\right) \left( 1-\alpha ^{2}r_{+}^{2}\right) }\left( E-\Omega _{H}J-%
\frac{qer_{+}}{\left( r_{+}^{2}+a^{2}\right) }\right) \right] .
\label{d12}
\end{equation}%
Note that the tunneling probability of scalar particles is the same
as in the case of Dirac particles \cite{RS}. This means that both
fermions and scalar particles emit at the same rate.

\section{The Acceleration Horizon}

We have seen in Section 2 that accelerating and rotating black holes
have acceleration horizons also, apart from the rotation horizons.
Here we find the tunneling probability at the acceleration horizon
$r_{\alpha }=1/\alpha $. The calculations for the probability at the
acceleration horizon proceeds in the same way as in the case of
outer horizon $r=r_{+}$. At the acceleration horizon $r_{\alpha
}=1/\alpha $, $W$ takes the form

\begin{eqnarray*}
W^{\prime 2}\left( r\right) &=&\frac{\left( r_{\alpha }^{2}+a^{2}\right) ^{2}%
}{4\alpha ^{2}\left[ r_{\alpha }^{2}-2Mr_{\alpha }+\left(
a^{2}+e^{2}+g^{2}\right) \right] ^{2}\left( r-r_{\alpha }\right)
^{2}}\left( E-\Omega _{H\alpha }J-\frac{qer_{\alpha }}{\left(
r_{\alpha
}^{2}+a^{2}\right) }\right) ^{2} \\
&&+\frac{\left( r_{\alpha }^{2}+a^{2}\cos ^{2}\theta \right)
}{2\alpha \left( 1-\cos \theta \right) ^{2}\left[ r_{\alpha
}^{2}-2Mr_{\alpha }+\left(
a^{2}+e^{2}+g^{2}\right) \right] \left( r-r_{\alpha }\right) }\left( \frac{%
\left( J-qA_{\phi }\left( r_{\alpha },\theta \right) \right)
^{2}}{K\left( r_{\alpha },\theta \right) }+m^{2}\right) .
\end{eqnarray*}%
Integrating as before we get
\begin{eqnarray*}
W_{\pm }\left( r\right) =\pm \int \frac{\left( r_{\alpha
}^{2}+a^{2}\right) }{2\alpha \left[ r_{\alpha }^{2}-2Mr_{\alpha
}+\left( a^{2}+e^{2}+g^{2}\right) \right] \left( r-r_{\alpha
}\right) }dr
\end{eqnarray*}%
$\times \sqrt{\left( E-\Omega _{H\alpha }J-\frac{qer_{\alpha
}}{\left( r_{\alpha }^{2}+a^{2}\right) }\right) ^{2}+\frac{2\alpha
\left( r_{\alpha }^{2}+a^{2}\cos ^{2}\theta \right) \left[ r_{\alpha
}^{2}-2Mr_{\alpha
}+\left( a^{2}+e^{2}+g^{2}\right) \right] \left( r-r_{\alpha }\right) }{%
\left( r_{\alpha }^{2}+a^{2}\right) ^{2}\left( 1-\cos \theta \right) ^{2}}%
\left( \frac{\left( J-qA_{\phi }\left( r_{\alpha },\theta \right)
\right) ^{2}}{K\left( r_{\alpha },\theta \right) }+m^{2}\right) }.$

\smallskip
Using the residue theory at the singularity $r=r_{\alpha }$ this
yields
\begin{equation}
W_{\pm }\left( r\right) =\pm \frac{\pi i\left( r_{\alpha
}^{2}+a^{2}\right) }{2\alpha \left[ r_{\alpha }^{2}-2Mr_{\alpha
}+\left(
a^{2}+e^{2}+g^{2}\right) \right] }\left( E-\Omega _{H\alpha }J-\frac{%
qer_{\alpha }}{\left( r_{\alpha }^{2}+a^{2}\right) }\right) .
\end{equation}%

Thus the tunneling probability of outgoing scalar particles
\begin{equation}
\Gamma =\exp \left[ -4 Im W_{+}\right] ,
\end{equation}%
becomes
\begin{equation}
\Gamma =\exp \left[ -\frac{2\pi \left( r_{\alpha }^{2}+a^{2}\right) }{\alpha %
\left[ r_{\alpha }^{2}-2Mr_{\alpha }+\left( a^{2}+e^{2}+g^{2}\right)
\right] }\left( E-\Omega _{H\alpha }J-\frac{qer_{\alpha }}{\left(
r_{\alpha }^{2}+a^{2}\right) }\right) \right] .
\end{equation}

\section{Conclusion}

Quantum tunneling of different particles from black hole horizons
has been studied in the literature. In this paper we have studied
the tunneling of uncharged and charged scalar particles from the
horizons of black holes with acceleration and rotation. For this
purpose we have solved the Klein-Gordon equation using the WKB
approximation and Hamilton-Jacobi method. We find that the tunneling
probability of scalar particles is the same as that of the Dirac
particles, which means that both Dirac particles and scalar
particles emit at the same rate. The correct Hawking temperature of
these black holes is also found using scalar particle tunneling
which is consistent with the literature.

\end{document}